\documentstyle[twoside,epsf]{article}

\catcode`\@=11
\long\def\@makefntext#1{
\protect\noindent \hbox to 3.2pt {\hskip-.9pt  
$^{{\eightrm\@thefnmark}}$\hfil}#1\hfill}		

\def\thefootnote{\fnsymbol{footnote}}
\def\@makefnmark{\hbox to 0pt{$^{\@thefnmark}$\hss}}	
	
\def\ps@myheadings{\let\@mkboth\@gobbletwo
\def\@oddhead{\hbox{}
\rightmark\hfil\eightrm\thepage}   
\def\@oddfoot{}\def\@evenhead{\eightrm\thepage\hfil
\leftmark\hbox{}}\def\@evenfoot{}
\def\sectionmark##1{}\def\subsectionmark##1{}}

\oddsidemargin=\evensidemargin
\renewcommand{\thefootnote}{\fnsymbol{footnote}}

\newcounter{sectionc}\newcounter{subsectionc}\newcounter{subsubsectionc}
\renewcommand{\section}[1] {\vspace{12pt}\addtocounter{sectionc}{1} 
\setcounter{subsectionc}{0}\setcounter{subsubsectionc}{0}\noindent 
	{\tenbf\thesectionc. #1}\par\vspace{5pt}}
\renewcommand{\subsection}[1] {\vspace{12pt}\addtocounter{subsectionc}{1} 
	\setcounter{subsubsectionc}{0}\noindent 
	{\bf\thesectionc.\thesubsectionc. {\kern1pt \bfit #1}}\par\vspace{5pt}}
\renewcommand{\subsubsection}[1] {\vspace{12pt}\addtocounter{subsubsectionc}{1}
	\noindent{\tenrm\thesectionc.\thesubsectionc.\thesubsubsectionc.
	{\kern1pt \tenit #1}}\par\vspace{5pt}}
\newcommand{\nonumsection}[1] {\vspace{12pt}\noindent{\tenbf #1}
	\par\vspace{5pt}}

\newcounter{appendixc}
\newcounter{subappendixc}[appendixc]
\newcounter{subsubappendixc}[subappendixc]
\renewcommand{\thesubappendixc}{\Alph{appendixc}.\arabic{subappendixc}}
\renewcommand{\thesubsubappendixc}
	{\Alph{appendixc}.\arabic{subappendixc}.\arabic{subsubappendixc}}

\renewcommand{\appendix}[1] {\vspace{12pt}
        \refstepcounter{appendixc}
        \setcounter{figure}{0}
        \setcounter{table}{0}
        \setcounter{lemma}{0}
        \setcounter{theorem}{0}
        \setcounter{corollary}{0}
        \setcounter{definition}{0}
        \setcounter{equation}{0}
        \renewcommand{\thefigure}{\Alph{appendixc}.\arabic{figure}}
        \renewcommand{\thetable}{\Alph{appendixc}.\arabic{table}}
        \renewcommand{\theappendixc}{\Alph{appendixc}}
        \renewcommand{\thelemma}{\Alph{appendixc}.\arabic{lemma}}
        \renewcommand{\thetheorem}{\Alph{appendixc}.\arabic{theorem}}
        \renewcommand{\thedefinition}{\Alph{appendixc}.\arabic{definition}}
        \renewcommand{\thecorollary}{\Alph{appendixc}.\arabic{corollary}}
        \renewcommand{\theequation}{\Alph{appendixc}.\arabic{equation}}
        \noindent{\tenbf Appendix \theappendixc #1}\par\vspace{5pt}}
\newcommand{\subappendix}[1] {\vspace{12pt}
        \refstepcounter{subappendixc}
        \noindent{\bf Appendix \thesubappendixc. {\kern1pt \bfit #1}}
	\par\vspace{5pt}}
\newcommand{\subsubappendix}[1] {\vspace{12pt}
        \refstepcounter{subsubappendixc}
        \noindent{\rm Appendix \thesubsubappendixc. {\kern1pt \tenit #1}}
	\par\vspace{5pt}}

\newcommand{\textlineskip}{\baselineskip=13pt}
\newcommand{\smalllineskip}{\baselineskip=10pt}

\def\eightcirc{
\begin{picture}(0,0)
\put(4.4,1.8){\circle{6.5}}
\end{picture}}
\def\eightcopyright{\eightcirc\kern2.7pt\hbox{\eightrm c}}

\def\abstracts#1#2#3{{
	\centering{\begin{minipage}{4.5in}\baselineskip=10pt\footnotesize
	\parindent=0pt #1\par 
	\parindent=15pt #2\par
	\parindent=15pt #3
	\end{minipage}}\par}} 

\newcommand{\bibit}{\nineit}

\renewenvironment{thebibliography}[1]
	{\frenchspacing
	 \ninerm\baselineskip=11pt
	 \begin{list}{\arabic{enumi}.}
	{\usecounter{enumi}\setlength{\parsep}{0pt}
	 \setlength{\leftmargin 12.7pt}{\rightmargin 0pt} 
	 \setlength{\itemsep}{0pt} \settowidth
	{\labelwidth}{#1.}\sloppy}}{\end{list}}

\newcounter{itemlistc}
\newcounter{romanlistc}
\newcounter{alphlistc}
\newcounter{arabiclistc}
\newenvironment{itemlist}
    	{\setcounter{itemlistc}{0}
	 \begin{list}{$\bullet$}
	{\usecounter{itemlistc}
	 \setlength{\parsep}{0pt}
	 \setlength{\itemsep}{0pt}}}{\end{list}}

\newenvironment{romanlist}
	{\setcounter{romanlistc}{0}
	 \begin{list}{$($\roman{romanlistc}$)$}
	{\usecounter{romanlistc}
	 \setlength{\parsep}{0pt}
	 \setlength{\itemsep}{0pt}}}{\end{list}}

\newenvironment{alphlist}
	{\setcounter{alphlistc}{0}
	 \begin{list}{$($\alph{alphlistc}$)$}
	{\usecounter{alphlistc}
	 \setlength{\parsep}{0pt}
	 \setlength{\itemsep}{0pt}}}{\end{list}}

\newcommand{\fcaption}[1]{
        \refstepcounter{figure}
        \setbox\@tempboxa = \hbox{\footnotesize Fig.~\thefigure. #1}
        \ifdim \wd\@tempboxa > 5in
           {\begin{center}
        \parbox{5in}{\footnotesize\smalllineskip Fig.~\thefigure. #1}
            \end{center}}
        \else
             {\begin{center}
             {\footnotesize Fig.~\thefigure. #1}
              \end{center}}
        \fi}

\newcommand{\tcaption}[1]{
        \refstepcounter{table}
        \setbox\@tempboxa = \hbox{\footnotesize Table~\thetable. #1}
        \ifdim \wd\@tempboxa > 5in
           {\begin{center}
        \parbox{5in}{\footnotesize\smalllineskip Table~\thetable. #1}
            \end{center}}
        \else
             {\begin{center}
             {\footnotesize Table~\thetable. #1}
              \end{center}}
        \fi}

\def\@citex[#1]#2{\if@filesw\immediate\write\@auxout
	{\string\citation{#2}}\fi
\def\@citea{}\@cite{\@for\@citeb:=#2\do
	{\@citea\def\@citea{,}\@ifundefined
	{b@\@citeb}{{\bf ?}\@warning
	{Citation `\@citeb' on page \thepage \space undefined}}
	{\csname b@\@citeb\endcsname}}}{#1}}

\newif\if@cghi
\def\cite{\@cghitrue\@ifnextchar [{\@tempswatrue
	\@citex}{\@tempswafalse\@citex[]}}
\def\citelow{\@cghifalse\@ifnextchar [{\@tempswatrue
	\@citex}{\@tempswafalse\@citex[]}}
\def\@cite#1#2{{$\null^{#1}$\if@tempswa\typeout
	{IJCGA warning: optional citation argument 
	ignored: `#2'} \fi}}

\def\pmb#1{\setbox0=\hbox{#1}
	\kern-.025em\copy0\kern-\wd0
	\kern.05em\copy0\kern-\wd0
	\kern-.025em\raise.0433em\box0}

\def\fnm#1{$^{\mbox{\scriptsize #1}}$}
\def\fnt#1#2{\footnotetext{\kern-.3em
	{$^{\mbox{\scriptsize #1}}$}{#2}}}

\def\fpage#1{\begingroup
\voffset=.3in
\thispagestyle{empty}\begin{table}[b]\centerline{\footnotesize #1}
	\end{table}\endgroup}

\def\runninghead#1#2{\pagestyle{myheadings}
\markboth{{\protect\footnotesize\it{\quad #1}}\hfill}
{\hfill{\protect\footnotesize\it{#2\quad}}}}
\font\tenrm=cmr10
\font\tenit=cmti10 
\font\tenbf=cmbx10
\font\bfit=cmbxti10 at 10pt
\font\ninerm=cmr9
\font\nineit=cmti9

\font\eightrm=cmr8

\def\qed{\hbox{${\vcenter{\vbox{			
   \hrule height 0.4pt\hbox{\vrule width 0.4pt height 6pt
   \kern5pt\vrule width 0.4pt}\hrule height 0.4pt}}}$}}

\renewcommand{\thefootnote}{\fnsymbol{footnote}}	

\makeatletter
\def\lesim{\mathrel{\mathpalette\gl@align<}}
\def\gtsim{\mathrel{\mathpalette\gl@align>}}
\def\gl@align#1#2{\lower.7ex\vbox{\baselineskip\z@skip\lineskip.2ex%
  \ialign{$\m@th#1\hfil##\hfil$\crcr#2\crcr\sim\crcr}}}
\makeatother

\begin{document}
\newcommand{\EQ}{\begin{equation}}
\newcommand{\EN}{\end{equation}}
\newcommand{\EQAN}{\begin{eqnarray*}}
\newcommand{\EQNN}{\end{eqnarray*}}
\newcommand{\EQA}{\begin{eqnarray}}
\newcommand{\EQN}{\end{eqnarray}}
\newcommand{\e}{{\rm e}}
\newcommand{\Sp}{{\rm Sp}}
\newcommand{\Tr}{{\rm Tr}}
\newcommand{\p}{\partial}
\newcommand{\EQAnn}{\begin{eqnarray*}}
\newcommand{\EQNnn}{\end{eqnarray*}}
\newcommand{\boldk}{\mbox{\boldmath $k$}}
\newcommand{\boldp}{\mbox{\boldmath $p$}}
\newcommand{\boldq}{\mbox{\boldmath $q$}}
\newcommand{\boldx}{\mbox{\boldmath $x$}}
\newcommand{\boldy}{\mbox{\boldmath $y$}}
\newcommand{\boldz}{\mbox{\boldmath $z$}}

\runninghead{Space-Time Uncertainty and Noncommutativity $\ldots$} {Space-Time Uncertainty and Noncommutativity  $\ldots$}

\normalsize\textlineskip
\thispagestyle{empty}
\setcounter{page}{1}



\fpage{1}
\centerline{\bf  Space-Time Uncertainty and Noncommutativity in String Theory}
\vspace*{0.37truein}
\centerline{\footnotesize TAMIAKI  YONEYA
}
\vspace*{0.015truein}
\centerline{\footnotesize\it  Institute of Physics, University of Tokyo
}
\baselineskip=10pt
\centerline{\footnotesize\it  
Komaba, Tokyo, 153-8902 Japan
}

\vspace*{0.21truein}
\abstracts{We analyze the nature of space-time 
nonlocality in string theory.   After giving a brief overview on 
the conjecture of the space-time uncertainty principle, a (semi-classical) reformulation of string 
quantum mechanics, in which the dynamics 
is represented by the noncommutativity 
between temporal and spatial coordinates, is outlined. 
The formalism is then compared to the space-time noncommutative field theories 
associated with nonzero electric $B$-fields.}{}{}


\vspace*{1pt}\textlineskip	
\section{Motivations}	
\vspace*{-0.5pt}
\noindent
What is string theory? This is a question we have been continually asking ourselves in exploring string theory as a 
hint towards the ultimate unified theory of 
all interactions including quantum gravity. 

One of the most characteristic features of string theory 
is the existence of a fundamental 
constant, string length $\ell_s \sim \sqrt{2\pi\alpha'}$, 
which sets a natural cutoff scale for the 
ultraviolet part of quantum fluctuations 
for particle fields associated with the 
spectrum of string states.  
This  implies 
that string theory must necessarily exhibit some  
nonlocality and/or certain fuzziness with respect to the 
short distance structure of space-time. 

From this point of 
view, it is quite remarkable that string theory gives  
a completely well-defined
 analytic S-matrix which essentially satisfies all the 
axioms for physically acceptable theory satisfying, 
at least perturbatively,  Lorentz invariance, (macro) causality and unitarity. In contrast to this, various
 past attempts toward nonlocal 
field theories failed to give sensible 
results, because of lack of  self-consistency 
or of suitable guiding 
principles for constructing nontrivially 
interacting theories. 
It thus seems an important task 
in uncovering its underlying principles 
 to characterize the nonlocality of string theory.  
Our attitude is that, given string theory, we should 
learn how to formulate the idea of fundamental 
length from the structure of string theory,  rather than 
postulating some arbitrary principles from scratch. 

Recent development   
on the connection of string theory with 
the external $B$-field to non-commutative 
geometric field theories might provide a good hint 
for pursuits in this direction.  However, we should keep in 
mind that the nature of nonlocality originated from the 
$B$-field is nothing to do with the extendedness of strings. 
In the present article, I would like to, first,  review briefly 
the old proposal of a `space-time uncertainty principle' 
as a possible general characterization of the 
space-time structure of string theory at short distances, 
and then to 
discuss some ideas toward a reformulation 
of string theory in such a way that the noncommutativity 
between space and time is manifest. I hope that the  
comparison of the nature of the latter noncommutativity 
to the one of the typical noncommutative 
field theories which have the algebra of space-time 
coordinates of the Moyal-type product associated with the 
electric B-field is useful for deepening our understanding of  string theory.  

\section{Space-Time Uncertainty}

The main idea for proposing the 
space-time uncertainty relation \cite{yo}\cite{yo1} 
comes from a simple analogy 
concerning the nature of string quantum mechanics. 
The crucial requirement of the ordinary 
string perturbation theory is the world-sheet conformal 
invariance.  Indeed,  most of the important merits 
of string theory as a possible unified theory are due to 
the conformal invariance. In particular, the 
elimination of the ultraviolet of divergence 
in the presence of gravity is essentially due to the 
modular invariance, which is the part of the 
conformal symmetry.  From the viewpoint 
of generic two-dimensional field theory, the 
conformal invariance forces us to choose a very 
particular class of all possible two-dimensional 
field theories, corresponding to  the fixed points 
of Wilsonian renormalization group.  
This is quite analogous to the imposition of 
Bohr-Sommerfeld quantization conditions to 
classical mechanics, in which the adiabatic invariance 
of action variables to be quantized can 
be regarded as a characterization of the 
quantization condition.  In the final formulation of 
quantum mechanics, the quantization condition 
was replaced by the more universal framework 
such as Hilbert space and operator algebra acting in it. 
This analogy suggests us the importance of reinterpreting the 
conformal invariance requirement by elevating it 
to a more universal form, which may ultimately be 
formulated in a way that does not depend
on perturbative methods.   

One of the crucial properties related 
to modular invariance is expressed as 
the `reciprocity relation' 
of the `extremal length'.   The  extremal length 
is a conformally invariant notion of length 
associated with families of curves on general 
Riemann surfaces.  If we consider some finite region $\Omega$ and 
a set $\Gamma$ of arcs on $\Omega$, the extremal 
length of $\Gamma$ is defined by 
$\lambda_{\Omega}(\Gamma) =
\sup_{\rho}{L(\Gamma, \rho)^2\over A(\Omega, \rho)}
$ with $L(\Gamma, \rho)=\inf_{\gamma\in \Gamma} L(\gamma, \rho) ,  \,
A(\Omega, \rho)=\int_{\Omega}\rho^2 dzd\overline{z} $ 
where $\rho$ is the possible metric function 
giving the length $L(\gamma, \rho)\equiv 
\int_{\gamma} \rho |dz|$ on $\Omega$ 
of a curve in $\Gamma$  
in the conformal gauge. 
Since any Riemann surface can be 
composed of a set of 
quadrilaterals pasted along the boundaries 
(with some twisting operations, in general), it is 
sufficient to consider the extremal length for an arbitrary  quadrilateral segment $\Omega$. Let the two pairs of 
opposite sides of $\Omega$ be $\alpha, \alpha'$  and 
$\beta, \beta'$. Take $\Gamma$ be the set of all 
connected set of arcs joining $\alpha$ and $\alpha'$. 
The set of arcs joining $\beta$ and $\beta'$ 
is called the conjugate set of arcs, denoted by $\Gamma^*$.  
We then have two extremal lengths, 
$\lambda_{\Omega}(\Gamma)$ and $\lambda_{\Omega}
(\Gamma^*)$. Then the reciprocity relation 
is that 
\EQ
\lambda_{\Omega}(\Gamma)\lambda_{\Omega}(\Gamma^*)=1 .
\label{reciprocity}
\EN
The simplest example is just the rectangle with the 
Euclidean sizes $a$ and $b$ in the Gauss plane. In 
this case, we can easily prove that $\lambda(\Gamma)=a/b,\quad \lambda(\Gamma^*)=b/a$.
For details, we refer the reader to a 
more extensive review \cite{yorev} and the mathematical references cited there. 

To see how the reciprocity of the extremal length 
reflects to target space-time, let us consider the 
Polyakov amplitude for the mapping from the 
rectangle on a Riemann surface to a 
rectangular region in space-time with the side lengths  
$A, B$ with the boundary condition $(0\le \xi_1\le a, 
0\le \xi_2\le 1)$
$
x^{\mu}(0, \xi_2)=x^{\mu}(a, \xi_2)=\delta^{\mu 2} B \xi_2/b,
x^{\mu}(\xi_1, 0)=x^{\mu}(\xi_1, b)=\delta^{\mu 1} A \xi_1/a.  $
Then the amplitude contains 
the factor 
\EQ
\exp\Bigl[
-{1\over \ell_s^2}\Bigl(
{A^2\over \lambda(\Gamma)} + {B^2\over \lambda(\Gamma^*)}
\Bigr)
\Bigr] ,
\label{abamp}
\EN
multiplied by a power-behaved prefactor. 
Thus the fluctuations of two space-time lengths $A$ and $B$ 
satisfy an `uncertainty relation' 
$\Delta A \Delta B \sim \ell_s^2$. 
For general and more complicated boundary conditions, 
it is not easy to establish a simple relation such 
as above  between the extremal lengths and the space-time
lengths,  since there are various ambiguities in defining 
 space-time lengths in terms of string variables. 
After all {\it only} legitimate observables 
allowed in string theory is the on-shell {\it S-matrix}. 
However, it seems natural to conjecture that the 
above relation sets a limitation, in some averaged sense,  on the measurability of 
the lengths in space-time in string theory, since 
conformal invariance must be valid to all orders 
of string perturbation theory and the random nature of 
boundaries generally contributes to further 
fuzziness on the space-time lengths. 
Note that this reciprocity relation 
exhibits one of the most important duality 
relations in string amplitudes between 
ultraviolet and infrared structures. 
Since in the Minkowski 
metric one of the lengths is always dominantly time-like, 
we propose the following uncertainty relation 
on the space-time lengths 
\EQ
\Delta T \Delta X  \gtsim \ell_s^2
\label{stu}
\EN
as a universal characterization of the 
short-distance space-time structure of string theory. 
This relation was originally proposed by the 
present author \cite{yo} in 1987 
independently of other proposals of similar nature, for example, the notion of  `minimal distance' \cite{gross}.  

The consistency of this `space-time uncertainty' relation 
with the high-energy behaviors of the 
perturbative string amplitudes was analyzed  
in ref. \cite{yorev} to which I refer the reader for 
details and relevant references.   We find that generically 
there are many instances where the above uncertainty 
relation is far from being saturated. However, so far all the 
known results seem to be
 consistent with the validity of the space-time 
uncertainty relation as an {\it inequality}. In particular, in
the  high-energy and high-momentum-transfer limit, 
both the temporal and spatial uncertainties increase 
linearly with respect to energy for fixed-genus amplitudes. 
However, the proportional constant decreases for higher genera 
and the well-known behavior  
$|A_{{\rm resum}}(s, \phi)| 
\sim \exp \big(-\sqrt{6\pi^2 sf(\phi)/\log s}\,\big)$ of the
Borel-summed amplitude \cite{menoo} is consistent with  the
saturation of the equality in (\ref{stu}) up to  some possible
logarithmic corrections that perhaps  depend on how to
precisely define the  space-time uncertainties.  This may be 
an indication that the relation (\ref{stu}) is  indeed valid
independently of string coupling $g_s$.  

A further support for the validity 
of the relation is its effectiveness for 
D-branes.  For example, the effective Yang-Mills 
theories for the low-velocity D-p-branes 
predict that the characteristic spatial ({\it transverse} to 
D-p-branes) and temporal 
scales of D-p-brane scattering oppositely scale  
with respect to the string coupling, namely 
as $\Delta X \sim g_s^{-1/(3-p)}\ell_s$ and $\Delta T \sim 
g_s^{1/(3-p)}$ for $p\ge 0$ and for $p\ne 3$.  Although the  case $p=3$  
is special in that the effective Yang-Mills theory 
is conformally invariant, the conformal transformation 
property is actually consistent with the space-time 
uncertainty relation as discussed in ref. \cite{jeyo}. 
 We can also derive these 
characteristic scales directly without recourse to 
the effective Yang-Mills theory. For example, the 
characteristic scales $\Delta X \sim g_s^{1/3}\ell_s, 
\Delta T\sim g_s^{-1/3}\ell_s$ of D-particle-D-particle scattering are a direct consequence \cite{liyo} of the 
space-time uncertainty relation and the 
ordinary quantum mechanical Heisenberg relation,  
given the fact that the mass of the D-particle is 
proportional to $1/g_s$.  
All these properties are natural from the viewpoint of 
open string theories where the relation (\ref{stu}) must 
be valid. 

Finally, let us discuss one important question. 
Since the relation (\ref{stu}) is independent of 
string coupling $g_s$, it seems at first sight that 
it does not take into account gravity. 
So what is its relation to the Planck scale 
which is the characteristic scale of quantum gravity?
In string theory, the existence of gravity
 \cite{yogra}$^,$ \cite{ss} can also 
be regarded as an important consequence of the 
world sheet conformal invariance. This is due to 
the possibility of deforming the background 
space-time by a linearized gravitational wave. 
However, in perturbation theory, the coupling strength of the gravitational wave is  
an independent parameter determined by the 
vacuum expectation value of dilaton. 
In this sense, the string coupling can not be 
a fundamental constant which appears in the 
universal nonperturbative property of string theory. 
Thus in oder to take into account the Planck length for the 
space-time uncertainty relation, we have to 
put that information by hand. Now we shall show that by 
combining the Planck scale with the space-time 
uncertainly relation, we can derive the M-theory 
scale without invoking D-branes or membranes. 

For that purpose, it is useful first to reinterpret the 
meaning of the Planck length using a similar language of 
the stringy 
space-time uncertainties, 
by considering the 
limitation of the notion of classical space-time as 
the background against the possible formation 
of virtual black holes in the short distance regime. 
If we want to probe the space-time structure 
in time direction to  order $\delta T$, 
the quantum mechanical uncertainty 
relation tells us that the 
uncertainty with respect to the energy 
of order $\delta E \sim 1/\delta T$ is necessarily 
induced. If we further require that the structure of the 
background space-time is not influenced 
appreciably by this amount of fluctuation, 
the spatial scale $\delta X$ to be probed can not be smaller than  the 
Schwarzschild radius associated with the 
energy fluctuation. Hence, $\delta X \gtsim (G_D\delta E)^{1/(D-3)}$ in $D$-space-time dimensions.  
This sets the relation for the characteristic 
gravitational scales in the form 
\EQ
\delta T (\delta X)^7 \gtsim g_s^{2}\ell_s^8. 
\label{bhrelation}
\EN
in $D=$10 dimensional string theory. This may be 
called the `black-hole uncertainty' relation. Note, 
however, that the nature 
of the scales $\delta T, \delta X$ is different from those in the relation (\ref{stu}).  The uncertainties in (\ref{bhrelation}) 
 only express  
limitations, for observers  
at asymptotic infinity,  with respect to 
 spatial and temporal resolutions,  below which 
the naive classical  space-time picture without 
the formation of microscopic black holes can no longer be applied. In contrast to this, the space-time uncertainty relation 
sets the more fundamental limitation 
below which the space-time geometry itself loses its meaning. 
 Then, the most important characteristic scale 
associated with the existence of gravitation 
in string theory corresponds to the point of their crossover. 
The critical crossover scales are then 
determined as 
\[
\Delta X_c \sim \delta X_c \sim g_s^{1/3}\ell_s , 
\quad 
\Delta T_c \sim \delta T_c \sim g_s^{-1/3}\ell_s  .
\]
It is quite remarkable that the spatial critical 
scale $\Delta X_c$ coincides with 
the M-theory scale.  Let us look at the 
Fig. 1 to appreciate the meaning of these scales. 
For $\Delta T < \Delta T_c$ 
there is no region where the fluctuation 
containing the microscopic black hole associated with 
quantum fluctuations is important in string theory,  
while, for $\Delta T > \Delta T_c$, 
there is a region where $(\Delta T)^{-1}\ell_s^2 < \Delta X < \Delta X_c$ 
is satisfied, and 
hence  black hole formation at the 
microscopic level becomes appreciable in string theory. 
The importance of this region increases as the string 
coupling grows larger.    
\begin{center}
\begin{figure}
\begin{picture}(180,160)
\put(80,0){\epsfxsize 180pt 
\epsfbox{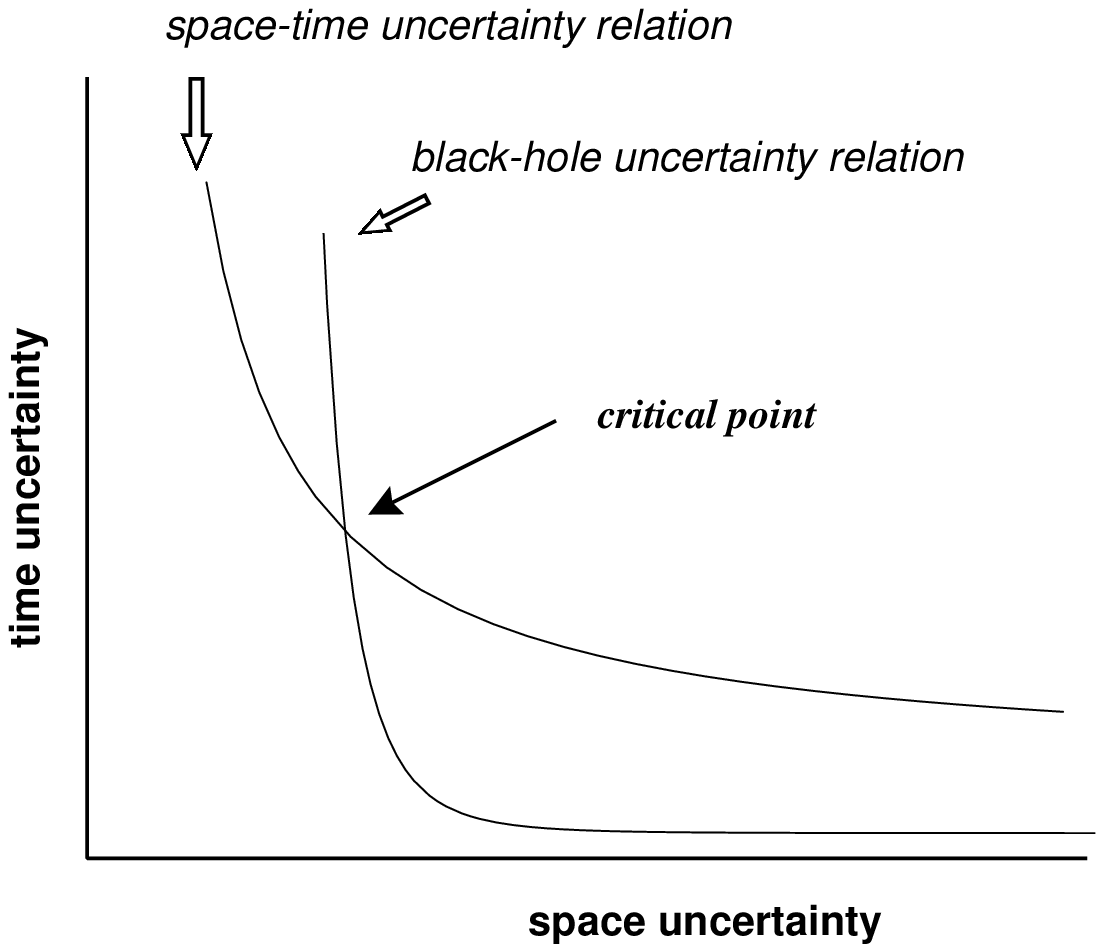}
}
\end{picture}
\caption{This diagram  
schematically shows the structure of the 
space-time uncertainty relation and the 
uncertainty relation associated with 
the Planck scale. The critical point is where the 
two relations meet. }
\label{Fig1} 
\end{figure}
\end{center}

\section{Space-Time Noncommutativity}

The validity of the fundamental uncertainty relation 
of the form (\ref{stu}) suggests the existence 
of certain noncommutative space-time 
structure that underlies string theory. 
Indeed, the expression (\ref{abamp}) is 
strongly reminiscent of the Wigner representation 
$\rho(p, q) \sim \exp \big[ 
-\big((p/\Delta p)^2 + (q/\Delta q)^2 \big)
\big]$ of the density matrix corresponding to 
the gaussian wave packet in particle quantum 
mechanics, suggesting that the space-time 
of string theory is something analogous to the 
classical phase space in particle quantum mechanics. 
However, the usual quantum mechanics of 
strings does not directly show any such noncommutativity 
 between space and time. 
This raises a question: Is there any alternative 
formulation of string quantum mechanics 
in which the space-time noncommutativity 
is manifest? 
In such a formulation \cite{yorev}, we expect that the 
world-sheet conformal symmetry 
would be translated into a quite 
different language. There might be, hopefully,  a chance 
of providing a hint toward some nonperturbative 
formulation of string theory.  

Let us start from the following version of the 
Nambu-Goto-Schild action:
\EQ
S_{ngs} =-\int_{\Sigma} d^2\xi\,  \Bigl\{{1\over e}
\Big[-{1 \over 2(4\pi \alpha')^2}(\epsilon^{ab}\partial_a X^{\mu}\partial_b  X^{\nu})^2\Big] + e
\Bigr\} .
\label{Schildaction}
\EN
We only consider the case of bosonic strings, expecting that the extension to 
superstrings will  
not cause any fundamental difficulty. 
By eliminating nonpropagating auxiliary field $e(\xi)$, 
the action reduces to the ordinary Nambu-Goto action. 
In this form, 
the conformal invariance of string theory is burried 
in the existence of  the standard Virasoro constraint 
${\cal P}^2 + {1\over 4\pi\alpha'}{\acute X}^2 = 0 , \quad 
{\cal P}\cdot {\acute X} =0$ 
which does {\it not} explicitly 
involve the world sheet auxiliary field and
the world-sheet metric.  It is important for later interpretation 
that the Hamiltonian constraint comes from the equation 
\EQ
 {1\over 4\pi \alpha'}\sqrt{-{1\over 2}(\epsilon^{ab}\partial_a X^{\mu}\partial_b  X^{\nu})^2}=e
\label{conformalconstraint}
\EN
for the auxiliary field $e$. 
Recall also that causality 
of string theory is embodied in the time-like 
nature of the area element $\epsilon^{ab}\partial_a X^{\mu}\partial_b  X^{\nu}$ in this formalism. 
Now in order to rewrite the action 
such that it becomes quadratic in the space-time 
coordinates, we introduce another auxiliary 
field $b_{\mu\nu}(\xi)$ which transforms as 
a world-sheet scalar and simultaneously as an 
antisymmetric tensor with respect to the 
space-time indices:
\EQ
S_{ngs2} =-\int_{\Sigma} d^2\xi\, \Bigl\{ {1\over 4\pi \alpha'}
\epsilon^{ab}\partial_a X^{\mu}\partial_b  X^{\nu}
b_{\mu\nu} + e\Big({1\over 2}b_{\mu\nu}^2 +1\Big) 
\Bigr\} .
\label{baction2}
\EN
Now the constraint 
(\ref{conformalconstraint}) is replaced by that
 for the new auxiliary field $b_{\mu\nu}$,  
\EQ
{1\over 2}b_{\mu\nu}^2 =-1 ,.
\label{bconstraint}
\EN 
The first auxiliary field $e$ only plays the 
role of Lagrange multiplier for this condition. Namely, the 
requirement of conformal invariance is essentially 
reinterpreted as the condition that the 
world-sheet $b$ field is time-like or `electric'.  

Let us consider the quantization of this 
action by regarding the $b$-field as an external field. 
Since the action is then first order with respect to 
the world-sheet time ($\tau$) derivative, the system has  
second class constraints 
$P_{\mu} = b_{\mu\nu}\partial_{\sigma} X^{\nu}/2\pi \alpha' , $ 
relating the components of the generalized 
coordinates and momenta directly.  
The Dirac bracket taking into account this constraint 
is given as 
\EQ
\Bigl\{
X^{\mu}(\sigma_1), 
\partial_{\sigma}b_{\nu\alpha}(\sigma_2)X^{\alpha}(\sigma_2)
\Bigr\}_D= 2\pi\alpha' \delta^{\mu}_{\nu}
\delta(\sigma_1 -\sigma_2) .
\label{diracbracket}
\EN
Remembering that the $b$-field is dominantly time-like 
by (\ref{bconstraint}), we see that the space $X^i(\xi)$ 
and the time $X^0(\xi)$  
become indeed noncommutative. In particular, 
the center-of-mass time 
$T\equiv (1/2\pi)\int_0^{2\pi} d\sigma X^0$ and 
the spatial extension $X$ defined by 
\EQ
X\equiv -\int d\sigma \, b_{0i}(\sigma) \partial_{\sigma}  X^i
\label{sextension}
\EN
satisfy ($2\pi\alpha'\rightarrow \ell_s^2$)
\EQ
\{T, X\}_D=\ell_s^2 .
\label{txcomm}
\EN
Here for simplicity we have assumed a closed string. 
That the expression (\ref{sextension}) can be 
adopted justifiably 
as the measure of spatial (longitudinal) extension of 
strings can be seen by remembering that in the semi-classical 
approximation the $b$-field is just proportional to 
the area element of the  world sheet of strings, 
$b^{\mu\nu} =-{1\over 4\pi \alpha'}
\epsilon^{ab}\partial_a X^{\mu}\partial_b  X^{\nu}/e$, 
which is derived by taking the variation of the action 
with respect to the $b$-field. 
Note that in this approximation the auxiliary 
$e$-field is determined by the normalization 
condition (\ref{conformalconstraint}) through  (\ref{bconstraint}). 
In this way, we have now reformulated the 
string mechanics, at least in the semi-classical 
approximation,  in such a way that the noncommutativity 
between spatial extension and time is manifest. This naturally 
conforms to the general property (\ref{stu}) of the space-time uncertainties, derived on the basis 
of the world-sheet conformal symmetry
 in the previous section, as 
it should be. For example, the constraint (\ref{bconstraint}) 
indeed replaces the role of conformal invariance. 

\vspace{0.2cm}
\noindent
{\it Remarks}

\noindent
1.   The nature of the 
noncommutativity discussed above is  close, at least formally, 
to that associated with the antisymmetric 
space-time external 
$B_{\mu\nu}(X)$ field in the presence of 
D-branes.  Note, however, 
an obvious difference that the present noncommutativity 
is intrinsic to the extendedness of strings and is nothing to do with the choice of the background of string theory. If we add the $B$-field background in considering D-branes,  
there arises an additional 
contribution to the noncommutativity, since 
the first term in the action 
is deformed as $b_{\mu\nu} \rightarrow b_{\mu\nu}+ 2\pi\alpha'
B_{\mu\nu}$.   Of course, when $B_{\mu\nu}$ is constant, it
affects only at the end points of open strings. 
Let us briefly treat the  open string boundary  
in the present formalism. 
If we allow free variations for $\delta X^{\mu}$ at the 
boundary without $B$-field along the D-brane world volume, 
the boundary condition is 
$\partial_{\tau}X^{\nu}b_{\mu\nu}=0$. 
In the sem-iclassical approximation we are using, 
we can set $b^{\mu\nu} =-{1\over 4\pi \alpha'}
\epsilon^{ab}\partial_a X^{\mu}\partial_b  X^{\nu}/e$. 
It is convenient here to choose the orthonormal world-sheet 
coordinate satisfying $\dot{X}\cdot X'=0$ and 
$\dot{X}^2+X'^2=0$ where the Lorentz contractions 
are done only over the directions along the 
D-brane world volume. Then, using 
(\ref{conformalconstraint}) the boundary condition becomes 
the usual Neumann condition $X'^{\mu}=0$.   
Now suppose we add a constant space-time $B$-field 
which corresponds to the additional boundary 
term $(1/2)\int_{\partial \Sigma} d\tau X^{\mu}(\tau)\partial_{\tau}
X^{\nu}(\tau)B_{\mu\nu}$. Then the boundary condition 
is $X'^{\mu}+ 2\pi\alpha'B^{\mu\nu}\dot{X}_{\nu}=0$. If the $B$-field is magnetic, it does not affect the 
time-like nature of the string coordinate at the boundary. 
However, if that is electric and the  
magnitude exceeds a critical value $1/2\pi\alpha'$ 
such that $b_{0i} + 2\pi\alpha' B_{0i}$ can vanish, 
the time-like nature of the open string boundary 
would be lost, leading to a violation of causality  
in the dynamics of open strings and D-branes. 
This can be seen as follows. Without losing generality 
we can assume that 
the only nonzero component of the $B$-field 
is $B_{01}=B$, providing  the direction 1 
is along the D-brane world volume. The boundary
condition together with  the coordinate condition leads 
to the relation  
\[
\dot{X}^2=-X'^2=(2\pi\alpha'B)^2((\dot{X}^1)^2-(\dot{X}^0)^2) .
\]
When $(2\pi\alpha'B)^2>1$, this implies that the vector 
$\dot{X}^{\mu}$ is space-like at the open-string boundary. 
Thus it is impossible \cite{seisuss} to decouple the string scale
     from  that of the noncommutativity associated to electric 
B-field. Namely, we can not define sensible field theory 
limits satisfying causality and unitarity 
using electric $B$ field \cite{gomes} without open-string 
degrees of freedom.  This seems to indicate that 
the feasibility of the space-time noncommutativity is
inextricably  connected to stringy nonlocality. 

\vspace{0.1cm}
\noindent
2. It is perhaps instructive to make a further 
comparison of the present formalism  
with the naive field theory model in which 
the noncommutativity between space and time is 
introduced explicitly.  A scalar field theory in a noncommutative 
space-time with space-time commutation relation 
$[x, t]=i\theta$ can be constructed by assuming the 
product of the fields are defined by the Moyal product 
\[
\phi (x)\ast \phi (x)=
e^{i{\theta\over 2}(\partial_{x_1} \partial_{t_2}-\partial_{t_1}\partial_{x_2})}
\phi (x_1)\phi (x_2)\Big|_{x_1=x_2=x, t_1=t_2=t} .
\]
For notational clarity, we consider only (1+1)-dimensional part of space-time. Then a 3-point interaction vertex takes the 
following form in the coordinate representation
\[
\hspace{-6.5cm}\int dxdt \, (\phi_1 \ast \phi_2 \ast \phi_3)(x,t)
\]
\EQ
\hspace{1.5cm}
=(\pi \theta)^{-2}\Big(\prod_{i=1}^3 \int dx_idt_i \Big)
\exp i\Big[\, 
{2\over \theta}\sum_{{\it cyclic}}\big(
x_i t_j -t_i x_j
\big)
\Big] 
\prod_{i=1}^3 \phi_i(x_i, t_i)  .
\label{3p}
\EN
The exponent ${2\over \theta}\sum_{{\it cyclic}}\big(
x_i t_j -t_i x_j
\big)$  in this expression is the formal 
analogue of the first term of the 
string action (\ref{baction2}) 
with the identification $\theta\sim \alpha'$. 
 Actually,  there is a crucial difference in that the exponential factor in field theory case directly 
leads to non-causal shifts of the time 
coordinates in proportion to the momenta  
of external lines as 
$
t_1-t_2 \sim \theta p_3 , \ldots.  
$
Hence the sign of the time shifts 
depends on the direction of momenta. 
In the case of strings,  
the connection between the 
external momenta and the shifts, if any,  of the 
time is not so direct as in the case of the 
simple Moyal product. For example, the 
center-of-mass spatial coordinates 
do not directly contribute to the noncommutativity 
in  (\ref{txcomm}). 
The dynamics of strings is completely local at each point 
of world sheet 
and hence the time-like nature of 
the area element ensures causality in the 
evolution of the system.  As discussed above,  causality 
is preserved as long as the external space-time 
electric $B$ field does not exceed the critical value. 

However, the above formal analogy prompts us to 
speculate a possibility of formulating the 
space-time uncertainty and noncommutativity 
as a certain kind of `deformation' 
from classical space-time geometry to 
quantum and stringy geometry.  It is a major challenge 
to find some unique characterization of such stringy 
deformation of space-time geometry. 

\vspace{0.1cm}
\noindent
3.  Another remark which might be useful in 
understanding the nature of the present reformulation is that 
the counterpart in particle theory of what we have done above is simply the momentum representation of the 
particle propagator.  We can start from the familiar 
particle action 
\[
L= -{1\over 2}\int d\tau \Big(
-{1\over e}({dx^{\mu}\over d\tau})^2 + e m^2
\Big). 
\]
By introducing another auxiliary field $p_{\mu}$  which is now 
a space-time vector corresponding to the 
line element of the world line, we can rewrite it 
as 
\[
L_p =\int d\tau 
\Big(
p_{\mu}{dx^{\mu}\over d\tau} -{1\over 2}e (p^2+m^2)
\Big) .
\]
Obviously, the role of the Virasoro condition in strings is now 
played by the mass-shell condition $p^2 + m^2=0 $ 
requiring that the momentum is time-like (or light-like 
when $m=0$). In the particle case, the action 
$\int d\tau\, p_{\mu}{dx^{\mu}\over d\tau}$ 
defines a usual Poisson structure. In analogy with this, 
the string action ${1\over 4\pi \alpha'}\int d^2\xi \, 
\epsilon^{ab}\partial_a X^{\mu}\partial_b  X^{\nu}
b_{\mu\nu}$ can be regarded as defining a 
generalized Poisson structure which is 
appropriate to strings\footnote{
After writing the previous review \cite{yorev},  the present
author came to know by reading reference \cite{kana} 
that a suggestion which is 
closely related to the present remark 
was first made by Nambu 
in \cite{nambu}. }. 
One natural possibility along this line would be to 
regard the auxiliary $b$-field as a sort of momentum 
variable corresponding to the area element of string 
world sheet.  Since, comparing to the particle case,  the 
string case has one additional dimension, this line of thought leads to the use of the so-called Nambu bracket. Such a possibility was 
indeed suggested in \cite{kana}. 
Unfortunately,  there seems to be
 no appropriate quantization procedure  
based on this interpretation.  The interpretation we have 
given in the present article 
by means of the ordinary Dirac bracket quantization 
seems to be the only viable possibility toward quantization. 
Our discussion, however, remains still at a very formal level. 
It is an open question whether the above formalism  
can lead to a new exactly 
calculable scheme in full-fledged quantum 
theory.  It is tempting to speculate 
a possibility of some tractable integral 
representation of string amplitudes 
where the ordinary moduli parameters of Riemann 
surfaces are integrated over. 
Instead of the moduli parameters, we should have 
some different integration variables corresponding 
to the `area momenta'.  
I leave such a possibility for the reader as an interesting new
direction in exploring string theory. 

\vspace{0.1cm}
\noindent
4.  
Another relevant question related to the more precise 
formulation  of the present approach 
is to discuss the curved background. 
A natural way of including the space-time 
metric is by introducing the viel-bein field 
$e_{\mu}^A(X)$ where $A$ is the local Lorentz 
index. We can assume that the auxiliary field 
$b_{AB}(\xi)$ now 
transforms as an antisymmetric tensor  at each local Lorentz frame on the 
space-time point $X^{\mu}(\xi)$ with the 
constraint $(1/2)b_{AB}^2=-1$. 
The area element is then written as 
$\epsilon^{ab}\partial_a X^{\mu}\partial_b  X^{\nu}
e_{\mu}^A(x(\xi)) e_{\nu}^B(x(\xi)) b_{AB}(\xi)$. 
The Dirac bracket relation is more 
complicated than the flat space, 
with the right hand side depending 
on the space-time coordinates. We expect that 
the requirement of consistent quantization 
would lead to the condition for the 
space-time background which should 
be equivalent to the familiar $\beta$-function 
condition of renormalization group. 
It is an important problem to work this out.  
Of course, once we could arrive at a satisfiable characterization 
of the deformed geometry as suggested above, such a 
property would be an evident consequence of the 
general formalism.   

\vspace{0.1cm} 
Even apart from further clarification 
and refinements of the ideas discussed here, 
there are innumerably many other remaining questions, 
such as the relevance of the space-time uncertainties  and 
noncommutativity for black-hole physics,   
the interpretation from the viewpoint of 
11 dimensional M-theory, consistency with S-and T-
dualities, interpretation of 
possible other scales than the M-theory scale,  
the role of supersymmetry, and so on.  
Some of these questions have been 
treated partially in ref. \cite{yorev}. 
Real answers to most of the deeper questions, however, must be left to the future.

\nonumsection{Acknowledgements}
I would like to thank the organizers of 
Strings 2000 conference for inviting me. 
The present work is supported in part by Grant-in-Aid for Scientific  Research (No. 12440060)  from the Ministry of  Education, Science and Culture.

\end{document}
Contributions to the {\it International Journal of Modern
Physics A} will be reproduced by photographing the author's
submitted typeset manuscript. It is therefore essential that the
manuscript be in its final form, and of good appearance because
it will be printed directly without any editing. The manuscript
should also be clean and unfolded. The copy should be evenly
printed on a high resolution printer (300 dots/inch or higher).
If typographical errors cannot be avoided, use cut and paste
methods to correct them. Smudged copy, pencil or ink text
corrections will not be accepted. Do not use cellophane or
transparent tape on the surface as this interferes with the
picture taken by the publisher's camera.
\pagebreak

\textheight=7.8truein
\setcounter{footnote}{0}
\renewcommand{\thefootnote}{\alph{footnote}}

\section{The Main Text}
\noindent
Contributions are to be in English. Authors are encouraged to
have their contribution checked for grammar. American spelling
should be used. Abbreviations are allowed but should be spelt
out in full when first used. Integers ten and below are to be
spelt out. Italicize foreign language phrases (e.g.~Latin,
French).

The text is to be typeset in 10 pt Times Roman, single spaced
with baselineskip of 13 pt. Text area (excluding running title)
is 5 inches (30 picas) across and 7.8 inches (47 picas) deep.
Final pagination and insertion of running titles will be done by
the publisher. Number each page of the manuscript lightly at the
bottom with a blue pencil. Reading copies of the paper can be
numbered using any legible means (typewritten or handwritten).

\section{Major Headings}
\noindent
Major headings should be typeset in boldface with the first
letter of important words capitalized.

\subsection{Sub-headings}
\noindent
Sub-headings should be typeset in boldface italic and capitalize
the first letter of the first word only. Section number to be in
boldface roman.

\subsubsection{Sub-subheadings}
\noindent
Typeset sub-subheadings in medium face italic and capitalize the
first letter of the first word only. Section number to be in
roman.

\subsection{Numbering and Spacing}
\noindent
Sections, sub-sections and sub-subsections are numbered in
Arabic.  Use double spacing before all section headings, and
single spacing after section headings. Flush left all paragraphs
that follow after section headings.

\subsection{Lists of items}
\noindent
Lists may be laid out with each item marked by a dot:
\begin{itemlist}
 \item item one,
 \item item two.
\end{itemlist}
Items may also be numbered in lowercase roman numerals:
\begin{romanlist}
\item item one
\item item two 
	\begin{alphlist}
	\item Lists within lists can be numbered with lowercase 
              roman letters,
	\item second item. 
	\end{alphlist}
\end{romanlist}
\newpage

\section{Equations}
\noindent
Displayed equations should be numbered consecutively in each
section, with the number set flush right and enclosed in
parentheses.
\begin{equation}
\mu(n, t) = {\sum^\infty_{i=1} 1(d_i < t, N(d_i) = n) \over
\int^t_{\sigma=0} 1(N(\sigma) = n)d\sigma}\,. \label{this}
\end{equation}

Equations should be referred to in abbreviated form,
e.g.~``Eq.~(\ref{this})'' or ``(2)''. In multiple-line
equations, the number should be given on the last line.

Displayed equations are to be centered on the page width.
Standard English letters like x are to appear as $x$
(italicized) in the text if they are used as mathematical
symbols. Punctuation marks are used at the end of equations as
if they appeared directly in the text.

\vspace*{12pt}
\noindent
{\bf Theorem~1:} Theorems, lemmas, etc. are to be numbered
consecutively in the paper. Use double spacing before and after
theorems, lemmas, etc.

\vspace*{12pt}
\noindent
{\bf Proof:} Proofs should end with \qed\,.

\section{Illustrations and Photographs}
\noindent
Figures are to be inserted in the text nearest their first
reference.  Original india ink drawings of glossy prints are
preferred. Please send one set of originals with copies. If the
author requires the publisher to reduce the figures, ensure that
the figures (including letterings and numbers) are large enough
to be clearly seen after reduction. If photographs are to be
used, only black and white ones are acceptable.

\begin{figure}[htbp]
\vspace*{13pt}
\centerline{\vbox{\hrule width 5cm height0.001pt}}
\vspace*{1.4truein}		
\centerline{\vbox{\hrule width 5cm height0.001pt}}
\vspace*{13pt}
\fcaption{Labeled tree {\footnotesize\it T}.}
\end{figure}

Figures are to be sequentially numbered in Arabic numerals. The
caption must be placed below the figure. Typeset in 8 pt Times
Roman with baselineskip of 10 pt. Use double spacing between a
caption and the text that follows immediately.

Previously published material must be accompanied by written
permission from the author and publisher.
\pagebreak

\section{Tables}
\noindent
Tables should be inserted in the text as close to the point of
reference as possible. Some space should be left above and below
the table.

Tables should be numbered sequentially in the text in Arabic
numerals. Captions are to be centralized above the tables.
Typeset tables and captions in 8 pt Times Roman with
baselineskip of 10 pt.

\begin{table}[htbp]
\tcaption{Number of tests for WFF triple NA = 5, or NA = 8.}
\centerline{\footnotesize NP}
\centerline{\footnotesize\smalllineskip
\begin{tabular}{l c c c c c}\\
\hline
{} &{} &3 &4 &8 &10\\
\hline
{} &\phantom03 &1200 &2000 &\phantom02500 &\phantom03000\\
NC &\phantom05 &2000 &2200 &\phantom02700 &\phantom03400\\
{} &\phantom08 &2500 &2700 &16000 &22000\\
{} &10 &3000 &3400 &22000 &28000\\
\hline\\
\end{tabular}}
\end{table}

If tables need to extend over to a second page, the continuation
of the table should be preceded by a caption, e.g.~``({\it Table
2. Continued}).''

\section{References}
\noindent
References in the text are to be numbered consecutively in
Arabic numerals, in the order of first appearance. They are to
be typed in superscripts after punctuation marks,
e.g.~``$\ldots$ in the statement.$^5$''.

\section{Footnotes}
\noindent
Footnotes should be numbered sequentially in superscript
lowercase Roman letters.\fnm{a}\fnt{a}{Footnotes should be
typeset in 8 pt Times Roman at the bottom of the page.}

\nonumsection{Acknowledgements}
\noindent
This section should come before the References. Funding
information may also be included here.

\nonumsection{References}
\noindent
References are to be listed in the order cited in the text. Use
the style shown in the following examples. For journal names,
use the standard abbreviations. Typeset references in 9 pt Times
Roman.

\appendix

\noindent
Appendices should be used only when absolutely necessary. They
should come after the References. If there is more than one
appendix, number them alphabetically. Number displayed equations
occurring in the Appendix in this way, e.g.~(\ref{that}), (A.2),
etc.
\begin{equation}
\mu(n, t) = {\sum^\infty_{i=1} 1(d_i < t, N(d_i) = n) \over
\int^t_{\sigma=0} 1(N(\sigma) = n)d\sigma}\,. \label{that}
\end{equation}
\end{document}